\documentclass{lcc}

\usepackage{url}

\usepackage{calc,xspace}
\usepackage[modulo]{lineno}  

\usepackage{amssymb} 
\usepackage{amsmath}  
\usepackage{amsbsy} 
\usepackage{amsfonts} 
\usepackage{euscript} 
\usepackage{graphicx} 
\usepackage{stmaryrd}

\usepackage{pstricks,pst-node,pst-text,pst-3d,pst-tree}

\newcounter{hours}\newcounter{minutes}

\newcounter{th}
\newenvironment{theorem}{\addtocounter{th}{1}\textbf{Theorem \thesection.\theth:}}{}  
\newcounter{ex}
\newenvironment{example}{\addtocounter{ex}{1}\textbf{Example \thesection.\theex:}}{}  
\newcounter{le}
\newenvironment{lemma}{\addtocounter{le}{1}\textbf{Lemma \thesection.\thele:}}{}  
\newcounter{de}
\newenvironment{definition}{\addtocounter{de}{1}\textbf{Definition \thesection.\thede:}}{}  
\newenvironment{proof}{\textbf{Proof:}}{}

\newcommand{\weight}{\omega}

\usepackage{calc,afterpage,xspace}
\usepackage{enumerate}
\usepackage{graphics}
\usepackage{amsmath,amssymb,stmaryrd}
\usepackage{color}
\usepackage{url}

\newcommand{\suc}{\textbf{S}}

\newcommand{\some}[3]{#1_{#2}, \cdots, #1_{#3}}
\newcommand{\many}[2]{\some{#1}{1}{#2}}

\newcommand{\Variables}{\mathcal{X}}
\newcommand{\Parameters}{\mathcal{P}}

\newcommand{\Functions}{\mathcal{F}}

\newcommand{\new}[1]{\texttt{new} \  #1}

\def\funone{\mbox{\texttt{f}}}

\newcommand{\getX}{\texttt{getX}}
\def\move{\mbox{\texttt{move}}}
\def\position{\mbox{\texttt{Position}}}

\def\funtwo{\mbox{\texttt{g}}}

\def\main{\mbox{\texttt{f}}}

\newcommand{\Constructors}{\mathcal{C}}

\def\conone{\mathbf{c}}

\def\skip{\texttt{skip}}
\def\return{\texttt{return}}
\def\command{\texttt{Cm}}
\def\class{\texttt{C}}
\def\Class{\texttt{Class}}
\def\main{\texttt{main}}

\newcommand{\ite}[3]{\texttt{if }#1 \texttt{ then } #2 \texttt{ else } #3}
\newcommand{\lop}[2]{\texttt{loop }#1 \left\{#2 \right\}}
\newcommand{\function}[3]{#1  \left\{#2 \ ;\ \return \ #3 ;\right\}}
\newcommand{\procedure}[2]{#1  \left\{#2 \right\}}
\newcommand{\constructor}[2]{#1  \left\{#2 \right\}}

\newcommand{\while}[2]{\texttt{while }#1 \left\{#2 \right\}}

\newcommand{\tstar}{\theta^*}

\newcommand{\dom}{\text{dom}}
\newcommand{\thetai}{I}
\newcommand{\tstari}{\thetai^*}

\newcommand{\bR}{\mathbb{R}}

\edef\eqone{e}

\edef\Attribut{A}
\edef\termone{e}

\newcommand{\zero}{\mbox{\textbf{0}}}

\newcommand{\add}{\mbox{\texttt{add}}}
\newcommand{\p}{\mbox{\texttt{p}}}

\newcommand{\un}{\textbf{1}}

\newcommand{\taille}[1]{|#1|}

\newcommand{\precF}{\geq_{\Functions }}

\newcommand{\sprecF}{>_{\Functions }}

\newdimen\proofrulebreadth \proofrulebreadth=.05em
\newdimen\proofdotseparation \proofdotseparation=1.25ex
\newdimen\proofrulebaseline \proofrulebaseline=2ex
\newcount\proofdotnumber \proofdotnumber=3
\let\then\relax
\def\hfi{\hskip0pt plus.0001fil}
\mathchardef\squigto="3A3B
\newif\ifinsideprooftree\insideprooftreefalse
\newif\ifonleftofproofrule\onleftofproofrulefalse
\newif\ifproofdots\proofdotsfalse
\newif\ifdoubleproof\doubleprooffalse
\let\wereinproofbit\relax
\newdimen\shortenproofleft
\newdimen\shortenproofright
\newdimen\proofbelowshift
\newbox\proofabove
\newbox\proofbelow
\newbox\proofrulename
\def\shiftproofbelow{\let\next\relax\afterassignment\setshiftproofbelow\dimen0 }
\def\shiftproofbelowneg{\def\next{\multiply\dimen0 by-1 }%
\afterassignment\setshiftproofbelow\dimen0 }
\def\setshiftproofbelow{\next\proofbelowshift=\dimen0 }
\def\setproofrulebreadth{\proofrulebreadth}

\def\prooftree{
\ifnum  \lastpenalty=1
\then   \unpenalty
\else   \onleftofproofrulefalse
\fi
\ifonleftofproofrule
\else   \ifinsideprooftree
        \then   \hskip.5em plus1fil
        \fi
\fi
\bgroup
\setbox\proofbelow=\hbox{}\setbox\proofrulename=\hbox{}%
\let\justifies\proofover\let\leadsto\proofoverdots\let\Justifies\proofoverdbl
\let\using\proofusing\let\[\prooftree
\ifinsideprooftree\let\]\endprooftree\fi
\proofdotsfalse\doubleprooffalse
\let\thickness\setproofrulebreadth
\let\shiftright\shiftproofbelow \let\shift\shiftproofbelow
\let\shiftleft\shiftproofbelowneg
\let\ifwasinsideprooftree\ifinsideprooftree
\insideprooftreetrue
\setbox\proofabove=\hbox\bgroup$\displaystyle 
\let\wereinproofbit\prooftree
\shortenproofleft=0pt \shortenproofright=0pt \proofbelowshift=0pt
\onleftofproofruletrue\penalty1
}

\def\eproofbit{
\ifx    \wereinproofbit\prooftree
\then   \ifcase \lastpenalty
        \then   \shortenproofright=0pt  
        \or     \unpenalty\hfil         
        \or     \unpenalty\unskip       
        \else   \shortenproofright=0pt  
        \fi
\fi
\global\dimen0=\shortenproofleft
\global\dimen1=\shortenproofright
\global\dimen2=\proofrulebreadth
\global\dimen3=\proofbelowshift
\global\dimen4=\proofdotseparation
\global\count255=\proofdotnumber
$\egroup  
\shortenproofleft=\dimen0
\shortenproofright=\dimen1
\proofrulebreadth=\dimen2
\proofbelowshift=\dimen3
\proofdotseparation=\dimen4
\proofdotnumber=\count255
}

\def\proofover{
\eproofbit 
\setbox\proofbelow=\hbox\bgroup 
\let\wereinproofbit\proofover
$\displaystyle
}%
\def\proofoverdbl{
\eproofbit 
\doubleprooftrue
\setbox\proofbelow=\hbox\bgroup 
\let\wereinproofbit\proofoverdbl
$\displaystyle
}%
\def\proofoverdots{
\eproofbit 
\proofdotstrue
\setbox\proofbelow=\hbox\bgroup 
\let\wereinproofbit\proofoverdots
$\displaystyle
}%
\def\proofusing{
\eproofbit 
\setbox\proofrulename=\hbox\bgroup 
\let\wereinproofbit\proofusing
\kern0.3em$
}

\def\endprooftree{
\eproofbit 
  \dimen5 =0pt
\dimen0=\wd\proofabove \advance\dimen0-\shortenproofleft
\advance\dimen0-\shortenproofright
\dimen1=.5\dimen0 \advance\dimen1-.5\wd\proofbelow
\dimen4=\dimen1
\advance\dimen1\proofbelowshift \advance\dimen4-\proofbelowshift
\ifdim  \dimen1<0pt
\then   \advance\shortenproofleft\dimen1
        \advance\dimen0-\dimen1
        \dimen1=0pt
        \ifdim  \shortenproofleft<0pt
        \then   \setbox\proofabove=\hbox{%
                        \kern-\shortenproofleft\unhbox\proofabove}%
                \shortenproofleft=0pt
        \fi
\fi
\ifdim  \dimen4<0pt
\then   \advance\shortenproofright\dimen4
        \advance\dimen0-\dimen4
        \dimen4=0pt
\fi
\ifdim  \shortenproofright<\wd\proofrulename
\then   \shortenproofright=\wd\proofrulename
\fi
\dimen2=\shortenproofleft \advance\dimen2 by\dimen1
\dimen3=\shortenproofright\advance\dimen3 by\dimen4
\ifproofdots
\then
        \dimen6=\shortenproofleft \advance\dimen6 .5\dimen0
        \setbox1=\vbox to\proofdotseparation{\vss\hbox{$\cdot$}\vss}%
        \setbox0=\hbox{%
                \advance\dimen6-.5\wd1
                \kern\dimen6
                $\vcenter to\proofdotnumber\proofdotseparation
                        {\leaders\box1\vfill}$%
                \unhbox\proofrulename}%
\else   \dimen6=\fontdimen22\the\textfont2 
        \dimen7=\dimen6
        \advance\dimen6by.5\proofrulebreadth
        \advance\dimen7by-.5\proofrulebreadth
        \setbox0=\hbox{%
                \kern\shortenproofleft
                \ifdoubleproof
                \then   \hbox to\dimen0{%
                        $\mathsurround0pt\mathord=\mkern-6mu%
                        \cleaders\hbox{$\mkern-2mu=\mkern-2mu$}\hfill
                        \mkern-6mu\mathord=$}%
                \else   \vrule height\dimen6 depth-\dimen7 width\dimen0
                \fi
                \unhbox\proofrulename}%
        \ht0=\dimen6 \dp0=-\dimen7
\fi
\let\doll\relax
\ifwasinsideprooftree
\then   \let\VBOX\vbox
\else   \ifmmode\else$\let\doll=$\fi
        \let\VBOX\vcenter
\fi
\VBOX   {\baselineskip\proofrulebaseline \lineskip.2ex
        \expandafter\lineskiplimit\ifproofdots0ex\else-0.6ex\fi
        \hbox   spread\dimen5   {\hfi\unhbox\proofabove\hfi}%
        \hbox{\box0}%
        \hbox   {\kern\dimen2 \box\proofbelow}}\doll%
\global\dimen2=\dimen2
\global\dimen3=\dimen3
\egroup 
\ifonleftofproofrule
\then   \shortenproofleft=\dimen2
\fi
\shortenproofright=\dimen3
\onleftofproofrulefalse
\ifinsideprooftree
\then   \hskip.5em plus 1fil \penalty2
\fi
}

\newcommand{\ninfer}[3]
     {\prooftree
          #1 
          \justifies #2
          \using #3
      \endprooftree}

\newcommand{\magique}{\diamond}

\begin{document}
\pagestyle{headings} 

\author{Jean-Yves Marion and Romain P\'echoux\\ \small Loria, Carte project, B.P. 239, 54506 Vand\oe  uvre-l\`es-Nancy Cedex, France, \\ \small and \'Ecole Nationale Sup\'erieure des Mines de Nancy, INPL, France. \\ \small \texttt{jean-yves.marion@loria.fr} \quad  \texttt{romain.pechoux@loria.fr}
}
            %


\title{Resource control of object-oriented programs} 
%
%
%
\maketitle

\begin{abstract} 
A sup-interpretation is a tool which provides an upper bound on the size of a value computed by some symbol of a program. Sup-interpretations have shown their interest to deal with the complexity of first order functional programs. For instance, they allow to characterize all the functions bitwise computable in \texttt{Alogtime}.
This paper is an attempt to adapt the framework of sup-interpretations to a fragment of oriented-object programs, including distinct encodings of numbers through the use of constructor symbols, loop and while constructs and non recursive methods with side effects.
We give a criterion, called brotherly criterion, which ensures that each brotherly program computes objects whose size is polynomially bounded by the inputs sizes.
\end{abstract}

\section{Introduction}
A sup-interpretation is a tool introduced in~\cite{MP06} in order to deal with the Implicit Computational Complexity of first order functional programs. Basically, the sup-interpretation of a first order functional program provides upper bounds on the size of any value computed by some function symbols of the program. The notion of sup-interpretation is a descendant of the notion of quasi-interpretation. See~\cite{BMM05b} for a survey on quasi-interpretations. It has been demonstrated in~\cite{MP07}, that the notion of polynomial sup-interpretation strictly generalizes the notion of polynomial quasi-interpretation. In other words, every polynomial quasi-interpretation is a polynomial sup-interpreta\-tion and there are programs which admit a sup-interpretation and no quasi-interpretation. As a consequence, sup-interpretation provides more intentionality than quasi-interpretation, i.e. it allows to capture the complexity of more algorithms. Such a flexibility is very interesting when we consider small complexity classes. For example, in~\cite{BMP06}, sup-interpretations allow to characterize all the functions bitwise computable in alternating logarithmic time. Another interesting consequence develloped in~\cite{MP06} consists in an application of the sup-interpretation tool to termination criteria such as the dependency pairs~\cite{AG00} or the size change principle~\cite{JonesSCP}.

The notion of quasi-interpretation has already been extended to Bytecode verification and to reactive programs. See for example~\cite{AmadioCDJ04,Amadio-Zilio04,DalzilioS:rbcvm}. Consequently, a major issue consists in the adaptation of the sup-interpretation analysis to imperative and object-oriented programs. We try to tackle this problem in this paper by enlarging the framework of sup-interpretations to a fragment of object-oriented programs without recursion. Our language is very similar to the language studied in~\cite{IPW01}. However, since we consider assignments, it is closer to a fragment of~\cite{DE99} where we add loop and while constructs. A consequence is that we have to control side effects. Our work is inspired by recent studies on the Implicit Computational Complexity of imperative programs~\cite{Niggl-matrices,JK06}. Contrarily to these seminal works, we work on polynomial algebra instead of matrix algebra. There are at least two reasons for such an approach. Firstly, the use of polynomials gives a clearest intuition and pushes aside a lot of technicalities. Secondly, polynomials give more flexibility in order to deal with method calls.  

The paper is organized as follows. After introducing our language and the notion of sup-interpretation of an object-oriented program, we give a criterion, called brotherly criterion, which ensures that each brotherly program computes objects whose size is polynomially bounded by the inputs sizes, even if the program is defined with function calls. To our knowledge, previous works on the implicit computational complexity of imperative programs did not support such a flexibility. Consequently, this criteria seems to be a great improvement on the study of the complexity of imperative programs. 

\section{Object-oriented Programs}\label{program}

\subsection{Syntax of programs} 
\label{sec:notations}
We consider object-oriented programs. 
Basically a program is composed by three sets of disjoint symbols $\Variables$, $\Parameters$ and $\Functions$ and a set $\Class \subseteq \Functions$. 
The set $\Variables$ represents the set of attributes. Throughout the following paper, we use capital letters $X,Y,Z,\ldots$ for attributes. 
The set $\Parameters$ is the set of parameters which are passed as arguments of a method. 
The symbols of $\Class$ are the class identifiers. They provide distinct data encodings such as the unary encoding, using the identifier $\suc$ for a class having one attribute and the identifier $\epsilon$ for a class without any attribute, or the binary encoding, using the class identifiers $\un$ and $\zero$. 
Each function symbol $\funone \in \Functions$ must be defined by one method of some class. 
A class $\class \in \Class$ is composed by attribute and method declarations, including a particular constructor method, which are described by the following grammar:

\label{def:global}      
$$
\begin{array}{lll}
\texttt{Attributes} &  \ni \Attribut & ::=  \texttt{var}\ X; \ | \ \texttt{var}\ X; \Attribut \\
\texttt{Expressions} &  \ni \termone & ::=  \ 
 x \ | \ X \ | \ X.\funone(\many{\termone}{n}) \ | \ \new{\class(\many{e}{n})} \\
\texttt{Commands} & \ni \texttt{Cm} & ::=  \ 
\skip \ | \ X:= \termone  \ | \ \command_1; \command_2 \ | \ \lop{X}{\command}\\  
&   & \ | \ \ite{e}{\command_1}{\command_2}  \ | \  \while{e}{\command} \\
\texttt{Methods} & \ni \texttt{M} & ::=  \function{\funone(x_1,...,x_n)}{\command}{ X} \\
\texttt{Class} & \ni \class & ::= \texttt{Class}\ \class \left\{ \Attribut \ ; \ \texttt{Cons} \ ; \ M_1 ; \ldots ; M_n;\right\} \\
 &  \main &::= \texttt{Class}\ \main \left\{ \Attribut \ ; \ \command \right\}
\end{array}
$$ 

where $X \in \Variables$, $x,\many{x}{n} \in  \Parameters$, $\funone \in \Functions$, $\many{e}{n} \in \texttt{Expressions}$ and $M_1 ; \ldots ; M_n \in \texttt{Methods}$. The method $\texttt{Cons}$ is a special constructor method of the shape $\constructor{\class(x_1,...,x_n)}{X_1:=x_1;\ldots;X_n:=x_n}$ which appears in each class $\class \in \Class$ whenever the class $\class$ has $n$ attributes $\many{X}{n}$. As a consequence, we have $\Class \subseteq \Functions $. This particular method can only be used in a command of the shape $X:=\new \ \class(\many{e}{n})$. 
All attributes appearing in the methods of a given class $\Class$ must belong to the attributes of this class. Finally, we define the main class $\main$ to be a special class defined by attributes declarations and commands.

We suppose that the attributes and methods of two distinct classes are pairwise distinct. 

For notational convenience, we sometimes refer to $\overline{e}$ as a sequence of expressions $\many{e}{n}$, whenever $n$ is clear from the context.

Given a program $\p$, we define a precedence $\precF$ on function symbols of $\Functions$. 
Set $\funone \precF \funtwo$ if the method defining $\funone$ is of the shape $\function{\funone(\overline{x})}{\command }{X}$ and the function symbol $\funtwo$ appears in $\command$. 
Take the reflexive and transitive closure of $\precF$, that we also note
$\precF$. It is not difficult to establish that $\precF$ is a preorder.
Lastly, we say that $\funone \sprecF \funtwo$ if $\funone \precF \funtwo$ 
and  $\funtwo \precF \funone$ does not hold. Intuitively, $\funone \sprecF \funtwo$  means that $\funone$ cannot call 
$\funtwo$. 
Throughout the following paper, we suppose that for each method of the shape $\function{\funone(\overline{x})}{\command }{X}$ and for each function symbol $\funtwo$ which occurs in $\command$, $\funone \sprecF \funtwo$, i.e. there is no recursive call in the program.

For each expression $e$ of a program, we suppose that function symbols $\funone \in \Functions - \Class$ appear only in the outermost position of an expression $e$. This restriction allows to deal with side effects in a clearest fashion. This is not a severe restriction since every program can be transformed efficiently in an equivalent program which does fit this requirement, by adding new attributes for the intermediate computations. For example, a command of the shape $X:=V.\funone(U.\funtwo(X))$ is transformed into $Y:=U.\funtwo(X);X:=V.\funone(Y)$ with $Y$ a fresh attribute. For simplicity, we suppose, that no function symbol appears in the expression $e$ of the commands $\ite{e}{\command_1}{\command_2}$ and $\while{e}{\command}$.

The attribute $X$ is not allowed to occur in the command $\command$ of an iteration of the shape $\lop{X}{\command}$. The attribute $X$ is not allowed to occur in a method of the shape $\function{\funone(\overline{x})}{\command'}{Y}$, if the function symbol $\funone$ appears in the command $\command$ of an iteration $\lop{X}{\command}$. In other words, the program is not allowed to read and to write the attribute $X$ during the execution of a $\lop{X}{\command}$.

\begin{example}\label{method}
Here is an example of a program of our language:
\begin{align*}
\Class \ \text{Position} \left\{ \right.& \\ 
&\texttt{var}\ X;\ \texttt{var}\ Y;\\
&\procedure{\position(x,y)}{X:=x ; Y:=y;}\\
&\function{\move(x,y)}{X:=X.\add(x) ; Y:=Y.\add(y)}{X}\\
&\function{\getX()}{\skip}{X}\\
\left.\right\}\quad \quad \quad \quad \quad \quad \quad &\\
\Class \ \text{main} \left\{ \right.& \\ 
&\texttt{var}\ W;\ \texttt{var}\ U;\ \texttt{var}\ V;\ \texttt{Var}\ Z;\\
&\command_1: V:=\new{\position(W,U)};\\
&\command_2: Z:=V.\move(W,W);\\
&\command_3: U:=V.\getX();\\
\left.\right\}\quad \quad \quad \quad \quad \quad \quad&
\end{align*}
where $\many{\command}{3}$ are labels used to refer to commands and $\add$ is a method which is not described in the program and which corresponds to the unary or binary addition depending on the kind of defined objects.
\end{example}

\subsection{Semantics}
The domain of computation is the set of objects defined inductively by:
$$\texttt{Objects}   \ni v  ::=  \ b \ | \ b(\many{v}{n}) \quad b \in \texttt{Class}$$
Given a $\main$ class having $n$ attributes $\many{X}{n}$, an object $v_i$ is stored in each $X_i$ at any time. 

A ground substitution $\sigma$ represents a store which consists in a total mapping from $\Variables $ to objects in $\texttt{Objects}$. Given a ground substitution $\sigma$ and an attribute $X$, the notation $\sigma \left\{X:=u\right\}$ means that the object stored in $X\sigma$ is updated to the object $u$ in $\sigma$. A parameter substitution $\beta$ is a total mapping from $\Parameters $ to objects in $\texttt{Objects}$. 
Given an expression $e$ and a ground substitution $\sigma$, we use the notation $\left\langle e, \sigma \right\rangle \rightarrow \left\langle u, \sigma' \right\rangle$ whether the expression $e$ evaluates to $u$ and the store $\sigma$ is updated to $\sigma'$. We use the notation $\left\langle \command, \sigma \right\rangle \rightarrow \left\langle \sigma' \right\rangle$, if $\sigma$ is updated to $\sigma'$ during the execution of the command $\command$. Given a program $\p$ of main class $\Class \texttt{ main } \left\{A ; \command \right\}$ and a store $\sigma$, $\p$ computes a store $\sigma'$ defined by $\left\langle \command, \sigma \right\rangle \rightarrow \left\langle \sigma'\right\rangle$.

The operational semantics of our language is inspired by the operational semantics of the java fragment given in~\cite{DE99} and is described in Figure~\ref{fig:cbv}. 


\begin{figure}[h]\label{fig:cbv}
\hrule
\begin{gather*}
\ninfer 
{ 
   } 
{ \left\langle \magique,\sigma\right\rangle \rightarrow \left\langle \magique\sigma, \sigma \right\rangle}  
{\magique \in \Variables \cup \Parameters }
\\[0.1cm] 
\ninfer 
{\left\langle \overline{e},\sigma\right\rangle\rightarrow \left\langle \overline{u},\sigma\right\rangle   \quad  \function{\funone(\overline{x})}{\command}{Y}\quad \exists \beta,\ \overline{x}\beta =\overline{u} \quad \left\langle \command\beta, \sigma \right\rangle \rightarrow \left\langle \sigma'\right\rangle 
  }
{ \left\langle X.\funone(\overline{e}),\sigma\right\rangle \rightarrow \left\langle Y\sigma' , \sigma ' \right\rangle}
{}
\\[0.1cm] 
\ninfer { \left\langle \overline{e},\sigma\right\rangle \rightarrow \left\langle \overline{u},\sigma\right\rangle} 
{ \left\langle \new \class(\overline{e}),\sigma\right\rangle \rightarrow \left\langle \class(\overline{u}),\sigma\right\rangle}
{\class \in \Class}
\\[0.1cm] 
\ninfer { \left\langle e,\sigma \right\rangle \rightarrow \left\langle u,\sigma'\right\rangle} 
{ \left\langle X:= e \right\rangle \rightarrow \left\langle \sigma\left\{X:=u\right\}\right\rangle}
{}
\\[0.1cm] 
\ninfer { } 
{\left\langle \skip, \sigma \right\rangle \rightarrow \left\langle \sigma \right\rangle}
{}
\\[0.1cm] 
\ninfer {\left\langle \command_1, \sigma \right\rangle \rightarrow \left\langle \sigma' \right\rangle \quad \left\langle \command_2, \sigma' \right\rangle \rightarrow \left\langle \sigma'' \right\rangle } 
{\left\langle \command_1;\command_2, \sigma \right\rangle \rightarrow \left\langle \sigma'' \right\rangle}
{}
\\[0.1cm] 
\ninfer {\left\langle e, \sigma \right\rangle \rightarrow \left\langle \underline{1},\sigma \right\rangle, \left\langle \underline{0},\sigma \right\rangle \text{ or } \left\langle u,\sigma \right\rangle} 
{\left\langle\ite{e}{\command_1}{\command_2},\sigma  \right\rangle \rightarrow \left\langle\command_1,\sigma  \right\rangle,\left\langle\command_2,\sigma  \right\rangle \texttt{ or } \left\langle\skip,\sigma  \right\rangle}
{\text{with } u>\underline{1}}
\\[0.1cm] 
\ninfer {  } 
{\left\langle\lop{X_i}{\command},\sigma \right\rangle \rightarrow \left\langle \command^{\taille{v_i}},\sigma\right\rangle}
{\texttt{with }\command^n =\command;\command^{n-1} \texttt{ and }\command^0=\skip}
\\[0.1cm] 
\ninfer {\left\langle e,\sigma\right\rangle \rightarrow \left\langle \underline{1},\sigma\right\rangle \text{ or } \left\langle \underline{1},\sigma\right\rangle} 
{\left\langle \while{e}{\command},\sigma \right\rangle \rightarrow \left\langle \command;\while{X}{\command},\sigma\right\rangle \texttt{ or } \left\langle \skip, \sigma \right\rangle}
{\texttt{with } u \neq \underline{1}}
\\[0.1cm] 
\end{gather*}
\caption{Call-by-value semantics} 
\label{fig:cbv}
\hrule
\end{figure}

If $\funone$ is defined by a method of the shape $\function{\funone(\overline{x})}{\command}{Y}$, then the evaluation of $X.\funone(\overline{u})$ is performed by first evaluating the body $\command \beta$, with $\beta$ a parameter substitution such that $\overline{x} \beta = \overline{u}$, in the context of the object $X$ and then returning the value stored in the attribute $Y$ of the object $X$.

The command $\skip$ does nothing. The command $X:= \termone$ assigns the value of $\termone$ to the attribute $X$. The command $X:= \new \class(\many{e}{n})$ first evaluates the expressions $\many{e}{n}$ to the objects $\many{v}{n}$, then, it creates a new object of the class $\class$ by assigning the value $\class(\many{v}{n})$ to the attribute $X$. The execution of $\command_1 ; \command_2$ corresponds to the sequential execution of $\command_1$ and $\command_2$. $\ite{b}{\command_1}{\command_2}$ executes the command $\command_1$, $\command_2$ or $\skip$ depending on whether the expression $b$ is respectively evaluated to an encoding of $1$, $0$ or another natural number. 
The size $\taille{v}$ of a value $v$ is defined to be the number of symbols of strictly positive arity in $v$. The command $\lop{X}{\command}$ executes $\taille{v}$ times the command $\command$ if $v$ is the value stored in $X$, i.e. $X\sigma=v$. Finally the command $\while{b}{\command}$ is evaluated to $\command ; \while{b}{\command}$ if $b$ is evaluated to an encoding of $1$ and to $\skip$ otherwise. 

\begin{example}
Consider the program of Example~\ref{method}. For each objects $u,v,w,z$ such that $\sigma=\left\{U:=u,V:=v,W:=w,Z:=z\right\}$, we have:
\begin{align*}
&\left\langle \new \ \position(W,U), \sigma \right\rangle \rightarrow \left\langle \position(w,u),\sigma \right\rangle\\
&\left\langle V:=\new \ \position(W,U), \sigma \right\rangle \rightarrow \left\langle \sigma\left\{V:=\position(w,u)\right\}\right\rangle
\end{align*}
Moreover if $v=\position(w,u)$ then:
\begin{align*}
&\left\langle V.\getX(), \sigma\right\rangle \rightarrow \left\langle w,\sigma\right\rangle\\
&\left\langle U:=V.\getX(),\sigma\right\rangle \rightarrow \left\langle \sigma\left\{U:=w \right\} \right\rangle
\end{align*}
\end{example}

\section{Sup-interpretations and weights}
\subsection{Assignments}
\begin{definition}\label{ass}
Given a class $\class$ having $n$ attributes $\many{X}{n}$, the assignment $\thetai$ of the class $\class$ is a mapping of domain $\dom(\thetai) \subseteq   \Functions$ which assigns a function $\thetai(\funone) : (\bR^+)^{m+1} \longmapsto \bR^+$ to every symbol $\funone \in \Functions-\Class$ of arity $m$, which corresponds to a method of the class $\class$, and which assigns a function $\thetai(\class) : (\bR^+)^{n} \longmapsto \bR^+$ to the constructor method of $\class$.

Given a program $\p$, the assignment $\thetai$ of $\p$ consists in the union of the assignments of each class $\class$ of $\Class$.

A program assignment $\thetai$ is defined over an expression $\eqone$ if each
symbol of $\Functions $ in $\eqone$ belongs to $\dom(\thetai)$. 
Suppose that the assignment $\thetai$ is defined over an expression $\eqone$,
The partial assignment of $\eqone$ w.r.t. $\thetai$, that we note $\tstari(\eqone)$ is the canonical extension of the assignment $\thetai$ defined as follows:
\begin{enumerate}
\item If $\magique$ is in $\Variables \cup \Parameters$, then $\tstari(\magique)=\magique$ 
\item If $\overline{e}$ is a sequence of expressions $\many{e}{k}$, $\tstari(\overline{e})=\tstari(e_1), \ldots ,\tstari(e_k)$.
%
%
\item If $\class$ is a symbol in $\Class$ of arity $m$ and $\many{\eqone}{m}$ are
  expressions,
  then, we have:  $$\tstari(\new \ \class(\many{e}{m}))=\thetai(\class)(\tstari(e_1),\ldots,\tstari(e_m))$$
\item If $\funone \in \Functions - \Class $ is a symbol of arity $m$ and $\many{\eqone}{m}$ are
  expressions,
  then, we have:  
\begin{align*}
\tstari(X.\funone(\many{e}{m}))&=\thetai(\funone)(\tstari(e_1),\ldots,\tstari(e_m),X)
\end{align*}
\end{enumerate}
\end{definition}

Notice that the assignment $\tstari(e)$ of an expression $e$ with $m$ parameters $\overline{x}$ occurring in a class $\class$ having $n$ attributes denotes a function from $(\bR^+)^{n+m} \to \bR^+$. Consequently, we use the notation $\tstari(e)(\many{X}{n},\overline{x})$ when we apply such a function.

\begin{definition}
Let $\textbf{Max-Poly}\left\{\bR^+ \right\}$ be the set of functions defined to be constant functions in $\bR^+$, projections, $\max$, $+$, $\times$ and closed by composition. Given a class with $n$ attributes, an assignment $\thetai$ is said to be polynomial if for every symbol $b$ of $\dom(\thetai)$, $\thetai(b)$ is a function of $\textbf{Max-Poly}\left\{\bR^+ \right\}$.
\end{definition}

\begin{definition}\label{add}
The assignment of a class symbol $\class \in \Class$ of arity $m>0$ is \emph{additive} if
$$  \thetai(\class)(\many{\magique}{m})  = \sum_{i=1}^m \magique_i + \alpha_{\class} \text{ where }
  \alpha_{\class} \geq 1$$
If the assignment of each class symbol of strictly positive arity is additive then the
assignment is additive. 
\end{definition}

\begin{definition}
The size of an expression $\eqone$ is noted $\taille{\eqone}$ and defined by
$\taille{\eqone} = 0$ if $\eqone$ is a $0$-ary symbol and $\taille{b(\eqone_1,\ldots,\eqone_m)} = 1 + \sum_i \taille{\eqone_i}$ 
if $\eqone=b(\eqone_1,\ldots,\eqone_m)$ with $m>0$.
\end{definition}

\begin{lemma}\label{valeur}
Given a program $\p$ having an additive assignment $\thetai$, there is a constant $\alpha$ such that
for each object $v \in \texttt{Objects}$, the following inequality is satisfied:
\begin{align*}
\taille{v} & \leq \tstari(v)  \leq  \alpha \times \taille{v} 
\end{align*}
\end{lemma}
\begin{proof}
Define $\alpha = \max_{\conone \in \Constructors}(\beta_{\conone})$ where $\beta_{\conone}$ is taken to be the constant $\alpha_{\conone}$ of definition~\ref{add} if $\conone$ is of strictly positive arity and $\beta_{\conone}$ is equal to the constant $\tstari(\conone)$ otherwise. The inequalities follow directly by induction on the size of a value.
\end{proof}

\subsection{Sup-interpretations}
\begin{definition}\label{SI} 
Given a program $\p$ of main class having $n$ attributes $X_1,\ldots,$ $X_n$, a sup-interpretation is an additive assignment $\theta$ of $\p$ which satisfies:
\begin{enumerate}
\item \label{SI1}
 The assignment $\theta$ is weakly monotonic. That is, for each symbol
 $b \in \dom(\theta)$, the function $\theta(b)$ satisfies:
$$\forall i,\ \magique_{i}\geq \magique'_{i} 
\Rightarrow 
\theta(b)(\ldots,\magique_i,\ldots) \geq \theta(b)(\ldots,\magique'_i,\ldots)
$$

\item  \label{SI3}
For each function symbol $\funone \in \dom(\theta)-\Class$ of arity $m$, for each $m$ tuple of objects $\overline{v}$, and for each store $\sigma$ if $\left\langle X_i.\funone(\overline{v}),\sigma \right\rangle \rightarrow \left\langle v,\sigma ' \right\rangle$ then
\begin{align*} 
\theta(\funone)(\tstar(\overline{v}), \tstar(X_i\sigma)) & \geq \max(\tstar(v),\tstar(X_i\sigma'))
\end{align*}
\end{enumerate}
\end{definition}

Intuitively, the sup-interpretation is a special interpretation of a symbol. Instead of yielding the symbol denotation, a
sup-interpretation of a function symbol provides an upper bound on the outputs sizes of the function denoted by the symbol. It is worth noticing that sup-interpretation is a complexity measure in the sense of Blum~\cite{B67}.

\begin{example}\label{3}
Suppose that the method $\add$ of Example~\ref{method} is defined over an encoding of unary numbers using two class constructor symbols $\suc$ and $\epsilon$ of respective arity $1$ and $0$. It admits the following additive and polynomial sup-interpretation $\theta(\add)(\magique_1,\magique_2)=\magique_1+\magique_2$, $\theta(\suc)(\magique)=\magique+1$ and $\theta(\epsilon)=0$. Indeed, this function is monotonic. For every unary number $\suc^v(\epsilon)$, we let the reader check that $\tstar(\suc^v(\epsilon))=\taille{\suc^v(\epsilon)}=v$. Moreover, for every unary number $\suc^v(\epsilon)$ and for every store $\sigma$ such that $X\sigma = \suc^{u}(\epsilon)$, with $\suc^{n+1}(\epsilon)=\suc(\suc^{n}(\epsilon))$ and $\suc^0(\epsilon)=\epsilon$, if $\left\langle X.\add(\suc^v(\epsilon)) , \sigma \right\rangle \rightarrow \left\langle \suc^{v + u}(\epsilon), \sigma\left\{X:=\suc^{v + u}(\epsilon)\right\} \right\rangle$, then:
\begin{align*}
\tstar(\add)(\tstar(\suc^v(\epsilon)),\tstar(\suc^u(\epsilon))) &= \tstar(\suc^u(\epsilon))+\tstar(\suc^v(\epsilon)) &&\text{By Dfn of $\theta$}\\
&= u+v &&\text{$\tstar(\suc^v(\epsilon))=v$}\\
&\geq \max(u+v,u) \\
&= \max( \tstar(\suc^{u+v}(\epsilon)),\tstar(\suc^{u}(\epsilon)))
\end{align*}
So that, Condition 2 of Definition~\ref{SI} is checked.
\end{example}

\begin{lemma}\label{deltaborne}
Given a program $\p$ of main class having $n$ attributes $X_1,\ldots,$ $X_n$ and having a sup-interpre\-tation $\theta$ defined over an expression $e$, then, for each parameter substitution $\beta$, $\tstar(e\beta)$ denotes a function from $(\bR^+)^n$ to $\bR^+$ which satisfies:

For each store $\sigma$, if $\left\langle e\beta, \sigma \right\rangle \rightarrow \left\langle v, \sigma'\right\rangle$ then
\begin{align*}
\tstar(e\beta)(\tstar(X_1\sigma),\ldots,\tstar(X_n\sigma)) &\geq \tstar(v)
\end{align*}
Moreover, if $e=X_i.\funone(\many{e}{n})$, we have:
\begin{align*}
\tstar(e\beta)(\tstar(X_1\sigma),\ldots,\tstar(X_n\sigma)) &\geq \tstar(X_i\sigma')
\end{align*}
\end{lemma}

\begin{example}
Consider the program of Example~\ref{method}. As demonstrated in Example~\ref{3}, $\theta(\add)(\magique_1,\magique_2)=\magique_1+\magique_2$, $\theta(\suc)(\magique)=\magique+1$ and $\theta(\epsilon)=0$ define a sup-interpretation for the method $\add$. 
The method $\move$ admits the following sup-interpretation $\theta(\move)(x,y,\magique)=x+y+\magique$ and $\theta(\position)(x,y)=x+y+1$. Indeed, $\theta(\move)$ and $\theta(\position)$ are monotonic and, for each store $\sigma=\left\{ U:=u,V:=\position(\suc^{u_1}(\epsilon),\suc^{u_2}(\epsilon)),W:=\suc^{w}(\epsilon),Z:=z\right\}$, we have: \begin{align*}
\left\langle V.\move(w,w),\sigma\right\rangle&\rightarrow \left\langle \suc^{u_1 + w}(\epsilon), \sigma\left\{V:=\position(\suc^{u_1 + w}(\epsilon),\suc^{u_2 + w}(\epsilon))\right\} \right\rangle
\end{align*}
Since $\tstar(\suc^n(\epsilon))=n=\taille{\suc^n(\epsilon)}$, we have to check Condition 2 of Definition~\ref{SI}:
\begin{align*}
&\tstar(V.\move(W\sigma,W\sigma))(\tstar(u,\position(\suc^{u_1}(\epsilon),\suc^{u_2}(\epsilon)),\suc^{w}(\epsilon),z)) \\
&= \theta(\move)(\tstar(W\sigma),\tstar(W\sigma),V)(\tstar(u,\position(\suc^{u_1}(\epsilon),\suc^{u_2}(\epsilon)),\suc^{w}(\epsilon),z))\\
&= \theta(\move)(\tstar(W\sigma),\tstar(W\sigma),\tstar(\position(\suc^{u_1}(\epsilon),\suc^{u_2}(\epsilon)))) \\
&=\tstar(W\sigma)+\tstar(W\sigma)+\tstar(\position(\suc^{u_1}(\epsilon),\suc^{u_2}(\epsilon)))\\
&=\tstar(\suc^{w}(\epsilon))+\tstar(\suc^{w}(\epsilon))+u_1+u_2+1\\
&\geq \max(w+u_1,2 \times w+u_1+u_2+1)\\
&=\max(\tstar(\suc^{u_1 + w}(\epsilon)),\tstar(\position(\suc^{u_1 + w}(\epsilon),\suc^{u_2 + w}(\epsilon))))
\end{align*}
\end{example}

\subsection{Weights}
Now, we are going to define the notion of weight which allows to control the size of the objects held by the attributes during loop and while iterations. Basically, a weight is a partial mapping over commands. 
The weights depend strongly on the considered command, so that we have to make the distinction between commands.

For that purpose, define the relation $\sqsubseteq$ over commands by $\command_1 \sqsubseteq \command$
\begin{itemize}
\item if there are $\command_2$ and $\command_3$ such that $\command= \command_2;\command_1;\command_3$,
\item $\command = \ite{e}{\command_2}{\command_3}$ and $\command_1 \sqsubseteq \command_2$ or $\command_1 \sqsubseteq \command_3$,
\item $\command = \lop{X}{\command_2}$ and $\command_1 \sqsubseteq \command_2$,
\item or $\command = \while{e}{\command_2}$ and $\command_1 \sqsubseteq \command_2$,
\end{itemize}
and its reflexive and transitive closure, that we also note $\sqsubseteq$. $\sqsubseteq$ defines a partial ordering over commands. The strict relation $\sqsubset$ is defined by $\command_1 \sqsubset \command$ if $\command_1 \sqsubseteq \command$ and $\command_1 \neq \command$

\begin{definition}
A command $\command$ is said to be:
\begin{itemize}
\item flat if there is no $\command_1$ of the shape $\command_1= \while{e}{\command_2}$ or $\command_1= \lop{X}{\command_2}$ such that $\command \sqsubseteq \command_2$.
\item minimum if there are no commands $\command_1$ and $\command_2$ and no expression $e$ such that $\command = \command_1;\command_2$ or $\command = \ite{e}{\command_1}{\command_2}$
\item whiled if there is a command $\command_1=\while{e}{\command_2}$ such that $\command \sqsubseteq \command_1$ or $\command_1 \sqsubseteq \command$
\item looped if if there is a command $\command_1=\lop{X}{\command_2}$ such that $\command \sqsubseteq \command_1$ and the command $\command_2$ is not whiled.
\end{itemize}
\end{definition}


\begin{definition}\label{wg} 
Given a program $\p$ having a main class with $n$ attributes, the weight of a command $\weight$ is a partial mapping. 
It assigns to:
\begin{itemize}
\item every flat, minimum and looped command $\command$, a total function $\omega_{\command}$ from $(\bR^+)^{n+1}$ to $\bR^+$ 
\item every flat, minimum and whiled command $\command$, a total function $\omega_{\command}$ from $(\bR^+)^n$ to $\bR^+$
\end{itemize}
which satisfy:
\begin{enumerate}
        \item $\weight_{\command}$ is weakly monotonic
$\forall i,\ \magique_i\geq \magique'_i \Rightarrow \weight_{\command}(\ldots,\magique_{i},\ldots) \geq \weight_{\command}(\ldots,\magique'_{i},\ldots)$
        \item $\weight_{\command}$ has the subterm property 
$\forall i,\ \forall \magique_i \in \bR^+ \ \weight_{\command}(\ldots,\magique_{i},\ldots) \geq \magique_{i}$
\end{enumerate}
A weight $\weight$ is polynomial if each $\weight_{\command}$ is a function of $\textbf{Max-Poly}\left\{\bR^+ \right\}$.
\end{definition}

\section{Criteria to control resources}
\subsection{Brotherly criterion}
The brotherly criterion gives constraints on weights and sup-interpretations in order to bound the size of the objects computed by the program by some polynomial in the size of the inputs.

\begin{definition}\label{brotherly}
A program having a main class with $n$ attributes $\many{X}{n}$ is \textbf{brotherly} if there are a polynomial sup-interpretation and a polynomial weight such that:
\begin{enumerate}
\item For every flat, minimum and looped command $\command$ of the main class, we have:\\
- For every expression of the shape $X_j.\funone(\many{e}{m})$ occurring in $\command$:
$$ \weight_{\command}(T+1,\many{X}{n})  \geq \weight_{\command}(T,X_1,\ldots,X_{j-1},\tstar(e)(\overline{X}),X_{j+1},\ldots,X_n)$$
- For every assignment $X_i:=e \sqsubseteq \command$, we have:
$$\weight_{\command}(T+1,\many{X}{n})  \geq \weight_{\command}(T,X_1,\ldots,X_{i-1},\tstar(e)(\overline{X}),X_{i+1},\ldots,X_n)$$
with $T$ is a fresh variable. 
\item For every flat, minimum and whiled command $\command$ of the main class, we have:\\
- For every expression of the shape $X_j.\funone(\many{e}{m})$ occurring in $\command$:
$$ \weight_{\command}(\many{X}{n})  \geq \weight_{\command}(X_1,\ldots,X_{j-1},\tstar(e)(\overline{X}),X_{j+1},\ldots,X_n)$$
- For every assignment $X_i:=e \sqsubseteq \command$, we have:
$$\weight_{\command}(\many{X}{n})  \geq \weight_{\command}(X_1,\ldots,X_{i-1},\tstar(e)(\overline{X}),X_{i+1},\ldots,X_n)$$
\end{enumerate}
\end{definition}

Intuitively, the first condition ensures that the size of the objects held by the attributes remains polynomially bounded. The fresh variable $T$ can be seen as a temporal factor which takes into account the number of iterations allowed in a loop. Such a number is polynomially bounded by the size of the objects held by the attributes. The second condition on whiled commands is very similar, however there is no more temporal factor, since we have no piece of information about the termination of a whiled command. \\
\begin{theorem}\label{borne}
Given a brotherly program $\p$ of main class $\Class \ \main \left\{A ; \command\right\}$ having $n$ attributes $\many{X}{n}$, there exists a polynomial $P$ such that for any store $\sigma$ if $\left\langle \command,\sigma\right\rangle \rightarrow \left\langle \sigma' \right\rangle$ then $$P(\taille{X_1\sigma},\ldots,\taille{X_n\sigma}) \geq \max_{i=1..n}(\taille{X_i\sigma'})$$
\end{theorem}

\begin{example}
The program of example~\ref{method} is brotherly since it admits a polynomial sup-interpretation $\theta$ and it has no looped and whiled command.
\end{example}

\begin{example} Consider the following program, over unary numbers:
\begin{align*}
\texttt{Class}\ \main &\left\{\right.\\
&\texttt{var}\ X_1;\\
&\texttt{var}\ X_2;\\
&\texttt{var}\ X_3;\\
&\texttt{loop } X_1 \left\{ \right.\\
& \quad \quad \quad \quad X_3 := X_3.\add(X_2);\\
 \left. \right\};&\\
\left.\right\}\quad \quad \quad \quad \quad&
\end{align*}
$\command = \lop{X_1}{X_3 := X_3.\add(X_2)}$ is the only minimum, flat and looped command. Applying the brotherly criterion, we have to find a polynomial weight $\weight_{\command}$ and a polynomial sup-interpretation $\theta$ such that:
$$\weight_{\command}(T+1,X_1,X_2,X_3)  \geq \weight_{\command}(T,X_1,X_2,\tstar(X_3.\add(X_2)))$$
with $T$ is a fresh variable.

Since $\theta(\add)(\magique_1,\magique_2)=\magique_1+\magique_2$ is a sup-interpretation of the method $\add$, previous inequality is satisfied by taking $\weight_{\command}(T,\magique_1,\magique_2,\magique_3)= T \times \magique_3 + \magique_1+\magique_2$ and the program is brotherly. 
\end{example}

\begin{scriptsize}
\bibliography{bib}
\bibliographystyle{plain}
\end{scriptsize}
\newpage

\appendix

\end{document}